# Bi-level Network Design for UAM Vertiport Allocation Using Activity-Based Transport Simulations

**Sebastian Brulin [1], Markus Olhofer [1]**

*1) Honda Research Institute Europe, Offenbach/Main, Germany*
*E-mail: sebastian.brulin@honda-ri.de*

**ABSTRACT**: The design or the optimization of transport systems is a difficult task. This is especially true in the case of the introduction of new transport modes in an existing system. The main reason is, that even small additions and changes result in the emergence of new travel patterns, likely resulting in an adaptation of the travel behavior of multiple other agents in the system. Here we consider the optimization of future Urban Air Mobility services under consideration of effects induced by the new mode to an existing system. We tackle this problem through a bi-level network design approach, in which the discrete decisions of the network design planner are optimized based on the evaluated dynamic demand of the user's mode choices. We solve the activity-based network design problem (AB-NDP) using a Genetic Algorithm on a multi-objective optimization problem while evaluating the dynamic demand with the large-scale Multi-Agent Transport Simulation (MATSim) framework. The proposed bi-level approach is compared against the results of a coverage approach using a static demand method. The bi-level study shows better results for expected UAM demand and total travel time savings across the transportation system. Due to its generic character, the demonstrated utilization of a bi-level method is applicable to other mobility service design questions and to other regions.

## 1. INTRODUCTION

Network design problems (NDP) have been the research subject for a long time (1, 2). They are relevant to many different domains, including transport, telecommunications, logistics, and economics, to name a few. Especially with the changing mobility landscape and the emergence of new multi-modal transport systems, the network design question is still very relevant. One complexity of network designs is the sometimes-counterintuitive effects that can arise. See Braess' Paradox, which demonstrates how the overall traffic flow can be slowed by adding more roads to a transport network.

Network design investigations have different degrees of inclusion of dynamics. This often determines how isolated or holistic the analysis is. One way to classify NDPs (3) is to distinguish between single-level approaches that investigate network designs without congestion effects, often used for investigating optimization methods (4,5) or minor adaptions to established networks. The second group of bilevel approaches incorporates transportation dynamics like congestion effects. The third group of multi-operator approaches, where our proposed method is set, tries to integrate the multi-stakeholders decision-making into the network design process, thereby allowing dynamics between the co-existing systems.

The authors of this study propose a bi-level optimization framework to solve the facility location for the example of the distribution of Urban Air Mobility (UAM) vertiports. The proposed framework is not limited to urban applications it can be applied to regional study, as shown later in this study. The bi-level investigation consists of two optimization processes linked and performed alternatingly. On the outer loop for the network optimization, a Genetic Algorithm (GA) is used with the Pareto optimal Non-Dominated Sorting Genetic Algorithm (NSGA-II) method to identify the best positions for UAM vertiports. The generated solutions are evaluated on a large-scale activity-based transport simulation (MATSim) within the inner loop, as shown in Fig. 1. In this inner loop a Co-Evolutionary Algorithm (CEA) based optimization is performed within MATSim to adapt the agents' activity plans reflecting the induced network changes. The agents aggregated activity patterns are afterward used for re-evaluating the network design. As a result of using a bi-level approach for the Activity-Based Network Design Problem (AD-NDP), it is possible to investigate the effect of the new transportation mode on all users. Thereby not only the reactions of the affected agents who use the UAM mode but also related effects, e.g., congestions due to network capacity or during rush hours and overall system travel times or resilience, can be incorporated and used as objectives in the design process. This integration enables the decision maker to compare optimized trade-off solutions for scenarios where mobility services must be designed together with multiple stakeholders.

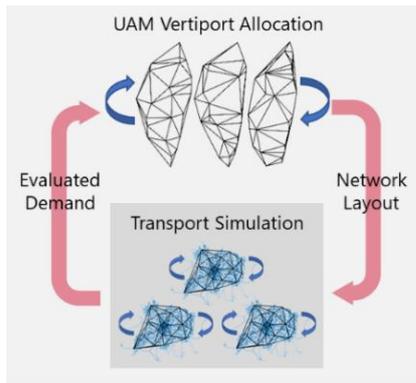

Fig. 2 Proposed bi-level optimization framework: UAM vertiport allocation on outer the layer and transport simulations on the inner layer.

The proposed framework is unique due to combining a bi-level approach for solving the network design problem in a multi-objective optimization approach with a large-scale activity-based transport simulation problem for solving the UAM vertiport placement problem.

## 2. APPROACH

The activity-based transport simulation method with its CEA optimization is described in 2.1, followed by the multi-objective optimization approach using a Genetic Algorithm to solve the Network Design Problem in 2.2.

### 2.1. Activity-Based Transport Simulation

The evaluation of the network designs is performed on an open-source platform for Multi-Agent Transport Simulation (MATSim) (6), which uses an activity-based approach that optimizes through a co-evolutionary approach the activity chains for a large number of agents in a 24h period. The MATSim framework and the scenario are open-source, and a description is available (6, 7). For our study, we used the available simulation of the region of Corsica[1] with a 1% representation of the population using 3400 agents. The agents can choose from the transportation modes car, walk, bike, train and bus to complete their daily activities, depending on individual and spatial accessibility. We extended the available transportation modes with a UAM system through the open-source extension MATSim-UAM from Bauhaus Luftfahrt (8). It allows for adding infrastructure and a transportation mode for an aerial mobility service. The vehicle configuration is based on data from the electrical Vertical Take-Off and Landing (eVTOL) database[2], using a mean set of parameters of 4 persons, 500 km range, and 250 km/h cruise speed.

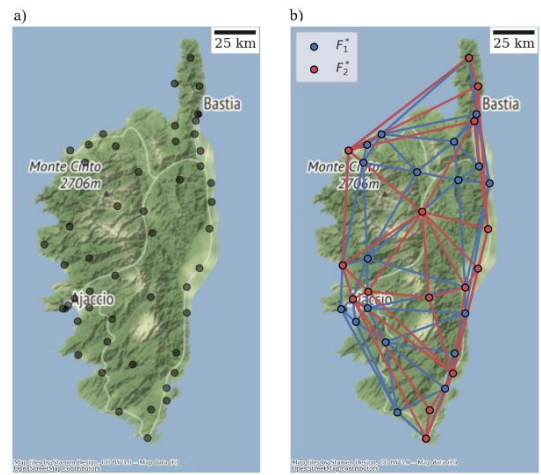

Fig. 1 a) Set of possible vertiport locations based on Corsica's population distribution; b) Multi-Objective Optimization results showing solution for UAM vertiport network with maximum demand ($f_1^*$) and minimum number of vertiports ($f_2^*$).

The set of possible vertiport locations is determined beforehand based on the residence locations of the agents. This reflects realistic constraints of the existing infrastructure, e.g., airports, helipads, or glider runways. For that, the population is separated into 50 clusters and based on a spatial mean center, possible vertiport locations are derived as shown in Fig. 2 a).

The optimized activity chains for each agent are explored within MATSim's activity-based simulation. This is accomplished through a CEA approach by mutating, e.g., transportation modes, start and end times, and activity orders. The generated activity chains are evaluated based on a utility function, which is, among other criteria, assessed on the travel time. Iteratively, each agent's activity chain is optimized until the transportation system converges, with the agents finding no better solutions under the network supply constraints. The transport system has then reached a stable equilibrium state.

### 2.2. Facility Location Problem

The following problem belongs to the facility location problems, a subclass of the NDPs, which targets distributing UAM vertiports by locating supply nodes in a transportation network to serve a nearby demand best. Comprehensive overviews of existing variants, including applications, are available from the literature (1, 3). In contrast to non-bi-level approaches where the static demand is known a priori through expert knowledge or simulations, our approach derives the demand individually for each network adaption from the bi-level activity-based simulation. Within the outer loop of the bi-level approach a multi-objective problem formulation is used. The first objective is targeting to maximize the overall UAM transportation demand

$$\max f_1 = \sum_{j \in N} d_j$$

with the UAM mode demand being $d_j$ for station $j$. The second objective minimizes the number of active UAM vertiports with $x_j$ being 1, if a facility is located at node j and 0 otherwise. The previously defined set of possible vertiports locations is $N$, such that $x_j \in \{0,1\}, j \in N$ and the objective being:

$$\min f_2 = \sum_{j \in N} x_j.$$

---

[1] github.com/matsim-scenarios/matsim-ile-de-france
[2] evtol.news/aircraft, accessed 20.01.2022.

The constraint for the number of active ports is thereby limited to be $P = 25$, where $P$ is defined as $\sum_{j \in N} x_j \leq P$.

For comparison, we chose a frequently used Heuristic Coverage Method (HCM) approach as baseline. The set of active facilities is optimized with the goal to maximize the number of agents within a predefined covering distance of an active vertiport. The implementation was taken from literature (2).

## 3. EVALUATION

Optimizing the facility location problem in the outer loop was performed on 50 generations with a population of 10 each. The co-evolutionary optimization within MATSim was performed for each of the 3400 agents. The convergence of trip distances and trip durations over all agents are shown in Fig. 3. Each simulation was performed ten times, during which the total times traveled decreased while the total distance traveled increased. This unintuitive observation comes due to the initial paths being generated from an A* Algorithm. Therefore they are the shortest paths but due to network limitations not the fastest. The initial paths still need to be optimized to be the fastest paths by considering the existing road capacities or dynamic events like congestion due to the transport choices of the other agents. Over ten iterations of MATSim's co-evolutionary optimization, the agents find improved paths that bring them faster to their goal but require longer travel distances. For our study, we chose that the travel distance saturation marks a satisfactory system stability to be used for solving our facility location problem.

The difference between the total distance traveled in the first and tenth iteration shows the Potential Travel Distance Saving

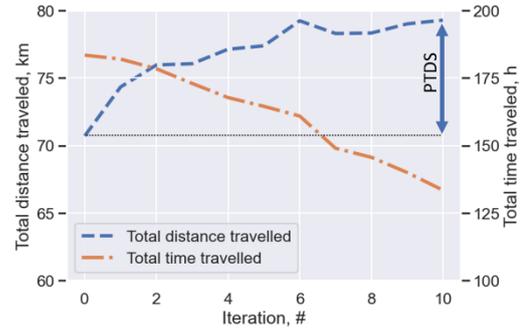

Fig. 3 Aggregated distance and time traveled of all agents with Potential Travel Distance Savings (PTDS).

(PTDS) of an existing network. In an ideal transport network, every agent could travel the shortest path with the shortest travel time, and the PTDS would be zero.

### 3.1. Results

The non-dominated solutions found by the NSGA-II are $f_1 \in [0,1]$ and $f_2 \in [15,25]$. $f_1$ is normalized by a maximum UAM demand of 187. The set of Pareto solutions can be found in Fig. 4, with a normalized $f_1$. The non-dominated solution for $f_1$ can be derived at the Pareto endpoint $F_1^* = (1.0, 25)$ with a vertiport network of 25 active ports. The non-dominated solution found for $f_2$ is $F_2^* = (0.4, 15)$ with a demand of 40 % and 15 active vertiports. Both found UAM network layouts for $F_1^*$ and $F_2^*$ are shown in Fig. 1 b). The network design for both networks shows that the overall reach from north-south and east-west is similar despite $F_1^*$ having ten additional active vertiports. The network layout $F_2^*$ suggests that the maximum demand with only 15 vertiports can be achieved by a network covering long distances at the coastal areas.

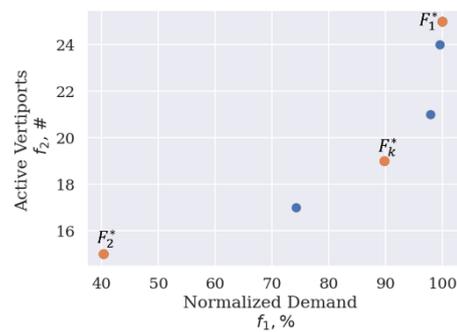

Fig. 4 Found solutions of the NSGA-II in the Pareto set.

A knee point solution within the Pareto set for the shown weighted approach would be $F_k^* = (0.9,19)$. This trade-off solution balances the minimization of vertiports and the demand and allows finding profitable service designs if cost and revenue structures are integrated into the parameterization. The corresponding network layout for $F_k^*$ is shown in Fig 5 b). The network maintains a similar north-south and east-west reach as $F_1^*$ and $F_2^*$ but with only 19 vertiports it has a different configuration in between.

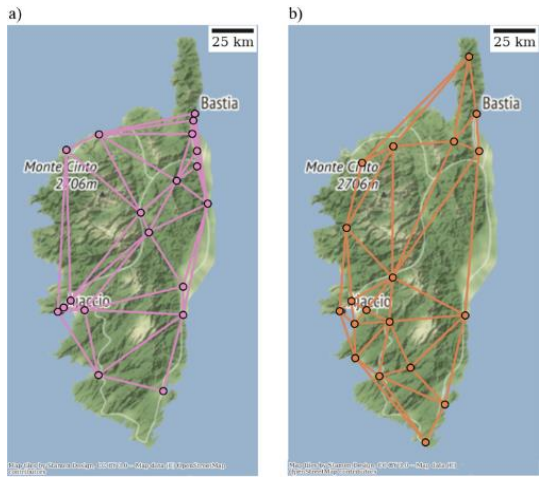

Fig. 5 UAM networks with 19 vertiports a) Heuristic Coverage Method (HCM) b) tradeoff solution of bi-level optimization $F_k^*$ (AB-NDP).

To compare the bi-level approach AB-NDP with the static facility location approach HCM, the HCM is applied on a single iteration of the simulation. For comparison, the number of vertiports is specified to the number found for the knee point solution $F_k^*(:,19)$.

The UAM demand shown in Table 1 is normalized with the maximum found UAM demand from $F_1^*$. The Total Travel Time Saving (TTTS) are the aggregated travel times across all modes. The travel times are normalized with the total travel time without a UAM transportation mode being available.

The solution from the AB-NDP approach shows a higher UAM demand than the HMC approach's solution by 19 %. The HMC UAM network in Fig. 5 a) indicates a more compact design than $F_k^*$ from Fig. 5 b). This is partly due to the HMC relying only on static locational information. In contrast, the AB-NDP has additional dynamic information about the activity locations of the agents, e.g., about their work, educational, or leisure areas, that are indirectly utilized within the bi-level framework. Additionally, the comparison of TTTS shows an improvement compared to a transport system without a UAM transportation mode. For the AB-NDP solution, the TTTS will increase by around 7.27 %, whereas for the HMC, it will only increase by 5.65 %. This shows the UAM transportation mode's effect beyond solving the vertiport location problem.

Table 1 UAM demand normalized with $F_1^*$; Total Travel Time normalized with total travel time w/o UAM transportation mode.

|  | Demand, % | TTTS, % |
|---|---|---|
| AB-NDP ($f_k^*$) | 89.98 | 7.27 |
| HMC | 70.97 | 5.65 |

The bi-level optimization framework AB-NDP shows better results than the HMC method for the investigated parameters and positively affects the overall transportation system.

## 4. CONCLUSION

A new bi-level AB-NDP approach was proposed to solve the facility location problem with a large-scale, open-source transport simulation to place UAM vertiports. The NDP was solved with an NSGA-II approach investigating the objectives of network size and UAM demand. The demand was evaluated through the activity-based transport simulation MATSim. Demand adaptions caused by transport network changes were incorporated into the design process.

An HCM without a bi-level coupling was used for comparison as a baseline. The results from the AB-NDP were superior for the investigated mode-specific UAM demand and transportation system-wide TTTS benchmark parameters.

Although adding a large-scale traffic simulation to a classical network design problem increases complexity by requiring additional expertise, it enables a holistic approach by incorporating co-existing system stakes into the mobility service planning process. Particularly for network design problems strongly influenced by the availability of other transportation systems, like sharing or swapping services, the bi-level approach presented can provide a solution well integrated into an existing multimodal transportation network.

## REFERENCES

(1) S. Hakimi, "Optimum locations of switching centers (…)", *Operations research*, 12(3), 450-459, 1964.

(2) R. Richard, and C. ReVelle, "The maximal covering location problem", Reg. sc. assoc. Vol. 32. No. 1., 1974.

(3) J. Chow, "Informed Urban transport systems: Classic and emerging mobility methods", *Elsevier*, 2018.

(4) T. Jatschka, T. Rodemann and G. Raidl, "Exploiting similar behavior of users in a cooperative optimization (...)", *Int Conf on Mach Learn, Opt, and DS*. Springer, Cham, 2019.

(5) T. Jatschka, G. Raidl and T. Rodemann, "A general cooperative optimization approach for distributing service points in mobility applications", Algorithms 14, 2021

(6) K. Axhausen, A. Horni, and K. Nagel, "The multi-agent transport simulation MATSim", *Ubiquity Press*, 2016.

(7) S. Hörl and M. Balać, "Open data travel demand synthesis", *Arbeitsberichte Verkehrs-und Raumplanung*, 1499, 2020.

(8) R. Rothfeld, M. Balac, K.O. Ploetner and C. Antoniou, "Agent-based simulation of urban air mobility", *Modeling and Simulation Technologies Conference* (p. 3891), 2018.